\documentclass[longbibliography,final,aps,prx,amsmath,amssymb,
			superscriptaddress,twocolumn]{revtex4-2}

\usepackage[T1]{fontenc}
\usepackage{lmodern}
\usepackage{blindtext}
\usepackage{amsfonts}
\usepackage{amsmath}
\usepackage{microtype}
\usepackage[usenames,dvipsnames]{xcolor}
\definecolor{shadecolor}{gray}{0.9}
\usepackage{graphicx}
\usepackage{overpic}
\usepackage{booktabs}
\usepackage{tabularx}
\usepackage{algorithm}
\usepackage{algpseudocode}
\usepackage{natbib}
\usepackage[colorlinks,linktoc=all]{hyperref}
\usepackage[all]{hypcap}
\definecolor{darkblue}{rgb}{0.0, 0.0, 0.55}
\definecolor{darkmidnightblue}{rgb}{0.0, 0.2, 0.4}
\definecolor{dukeblue}{rgb}{0.0, 0.0, 0.61}
\definecolor{zaffre}{rgb}{0.0, 0.08, 0.66}
\hypersetup{citecolor=zaffre}
\hypersetup{linkcolor=zaffre}
\hypersetup{urlcolor=zaffre}
\usepackage[nameinlink,capitalize]{cleveref}
\usepackage{todonotes}
\usepackage{mathtools}
\usepackage{amsthm}

\newcommand{\be}{\begin{equation}}
\newcommand{\ee}{\end{equation}}
\newcommand{\bea}{\begin{eqnarray}}
\newcommand{\eea}{\end{eqnarray}}

\newcommand{\f}[2]{\frac{#1}{#2}}

\newcommand{\bup}[1]{\left(#1\right)}

\usepackage{printlen}

\newcommand{\Graph}{G}
\newcommand{\Nodes}{V}
\newcommand{\Edges}{E}
\newcommand{\Nnode}{{|V|}}

\newcommand{\NetMass}{S}
\newcommand{\Inc}{B}

\newcommand{\Inode}{v}

\newcommand{\Inodethree}{u}

\newcommand{\Iedge}{e}

\newcommand{\Flux}{F}

\newcommand{\Cond}{\mu}

\newcommand{\Press}{p}
\newcommand{\Lap}{L}

\theoremstyle{definition}

\newtheorem{propo}{Proposition}

\DeclarePairedDelimiter\floor{\lfloor}{\rfloor}

\crefrangeformat{equation}{eqs. #3(#1)#4-#5(#2)#6}
\crefformat{equation}{Eq.~#2(#1)#3}
\crefname{eqnarray}{eq.}{eqs.}
\Crefname{eqnarray}{Eq.}{Eq.}
\crefname{figure}{fig.}{figs.}
\Crefname{figure}{Fig.}{Figs.}

\usepackage[caption=false]{subfig}
\usepackage[normalem]{ulem} 

\newcommand{\kl}{Kirchhoff's}

\begin{document}

\title{Infrastructure adaptation and emergence of loops in network routing with time-dependent loads}
\author{Alessandro Lonardi}
\email{alessandro.lonardi@tuebingen.mpg.de}
\affiliation{Max Planck Institute for Intelligent Systems, Cyber Valley, T{\"u}bingen 72076, Germany}
\author{Enrico Facca}
\affiliation{Laboratoire Paul Painlevé, UMR No. 8524, CNRS,  Inria, Université Lille, 59000 Lille, France}
\author{Mario Putti}
\affiliation{Department of Mathematics ``Tullio Levi-Civita'', University of Padua, Via Trieste 63, 35131 Padua, Italy}
\author{Caterina De Bacco}
\email{caterina.debacco@tuebingen.mpg.de}
\affiliation{Max Planck Institute for Intelligent Systems, Cyber Valley, T{\"u}bingen 72076, Germany}

\begin{abstract}
Network routing approaches are widely used to study the evolution in time of self-adapting systems.  However, few advances have been made for problems where adaptation is governed by time-dependent inputs. In this work we study a dynamical systems where the edge conductivities of a network are regulated by time-varying mass loads injected on nodes.  Motivated by empirical observations, we assume that conductivities adapt slowly with respect to the characteristic time of the loads. Furthermore, assuming the loads to be periodic,  we derive a dynamics where the evolution of the system is controlled by a matrix obtained with the Fourier coefficients of the input loads. Remarkably, we find a sufficient condition on these  coefficients that determines when the resulting network topologies are trees. We show an example of  this on the Bordeaux bus network where we tune the input loads to interpolate between loopy and tree topologies. We validate our model on several synthetic networks and provide an expression for long-time solutions of the original conductivities.
\end{abstract}

\maketitle

\section{Introduction}

Optimized transport of resources is a pivotal contributing factor in determining the structural evolution of real-world networks. Archetypes for self-organizing systems that ramify into networks in order to optimize energy expenditure rates are xylem conduits in leaves \cite{ronellenfitsch2016global, ronellenfitsch2019phenotypes, katifori2010damage, xia2007formation}, river basins \cite{sinclair1996, rinaldo1992minimum, rinaldo1993self, sun1994minimum, konkol2021interplay}, and slime molds \cite{Tero439, tero2008flow, tero2006computation, tero2007mathematical, yamada2000intelligence, bonifaci2013short, bonifaci2017revised, baptista2020network, baptista2021principled,baptista2021convergence}. These formations are not only restricted to the natural realm but can also be generated by anthropogenic processes. A prominent example is that of transportation networks such as railway and metro systems, which are designed to jointly optimize traffic overload and infrastructural cost \cite{adinoyi2021optimal, bonifaci2021physarum, yeung2013physics}.

Typically, optimal transport of mass in networks is set as a minimization problem where resources moving through the edges have to satisfy a set of constraints, e.g. conservation of mass, while minimizing a suitable transportation cost \cite{bohn2007structure, katifori2010damage, corson2010fluctuations, hu2012biological, ronellenfitsch2016global, kirkegaard2020optimal, lonardi2021designing, lonardi2021multicommodity, maritan, xia2003optimal,baptista2021convergence}. 

Several efficient methods have been proposed to solve this problem. A popular approach is that of message-passing algorithms \cite{mezard2009information}, where sources of mass are matched in sender-receiver pairs, and messages encode mass transfer between them \cite{yeung2012competition, yeung2013networking, yeung2013physics, altarelli2015edge, de2014shortest, xu2021scalable}. Promising results have also been obtained with optimal transport theory \cite{hu2013adaptation, ronellenfitsch2019phenotypes, lonardi2021designing, adinoyi2021optimal, bonifaci2012physarum, bonifaci2013short, facca2016towards, facca2019numerics, facca2020branching, facca2020physarum}, the approach we consider in this work. The general idea behind this method is to describe the transport of mass as a process being regulated by edge capacities, quantities evolving with a dynamical system to allocate mass fluxes.

Despite their usage in modeling transportation problems across domains, a common drawback of all these methods is to consider only stationary loads, i.e. resources that are injected and travel through the network do not change with time. This assumption may not be valid in certain scenarios. For instance, blood vessels are known for adapting their structure continuously to meet changing metabolic demands \cite{folkow1963studies, granger1982role, widmer2007application, hu2012biological}. Similarly, passengers in transportation systems enter stations with hourly, weekly, and seasonal time-varying rates \cite{londondata}. 

A viable approach to model these systems is to control the network evolution considering an ensemble average of the stress generated by the loads \cite{hu2013adaptation, corson2010fluctuations, hu2012biological}. This relies on assuming stationary loads on nodes but with their positions varied stochastically. The ensemble average over the loads' locations is then computed as a proxy of a system with loads of fixed locations but time-varying amounts. This technique has also been employed to study network resilience to edge cutting \cite{katifori2010damage} or for routing problems with spatially correlated loads \cite{ronellenfitsch2019phenotypes}.  

Remarkably, adding stochasticity in the loads may lead to the emergence of loops in the resulting optimal networks topologies \cite{hu2013adaptation, corson2010fluctuations, hu2012biological, katifori2010damage, ronellenfitsch2019phenotypes}. This result is complementary to the hierarchical formation of trees since loops provide alternative routes to accommodate fluctuations, or guarantee robustness against broken links. 
Recently,  loops formation has also been observed in multicommodity setups \cite{lonardi2021designing,lonardi2021multicommodity}, where the loads are deterministic inputs of the problem. In this case loop generation is a consequence of having different commodities interacting in a unique shared infrastructure. 

In all these works, the time-varying character of the transport network loads is neglected because the main problem variables are taken on average.  Here we develop a model that considers the explicit time dependence of the mass inflows and investigate, both analytically and numerically, the long-term behavior of time-varying transport networks. This allows us to show that it is not the process uncertainty, inherent in any stochastic framework, but the non stationarity of the loads that promotes loops, which is fundamentally different from what can be concluded using stochastic formulations. 

In particular, we generalize the routing problem in the work of Facca \textit{et al.} \cite{facca2020branching} by considering periodic mass loads on nodes. We postulate an analytical relationship connecting physical quantities as the edge conductivities and the coefficients of the Fourier series expansion of the loads.  We then define a dynamics that rapidly converges to the long-run average solutions of the original dynamics. 

Our model relies on the idea of distinguishing slow-varying variables from fast ones. The first are capacities of edges that are regulated by the Fourier coefficients of the forcing; the second are, for example, loads of passengers entering and exiting network nodes. The physical intuition is that, while fluxes of passengers in a transportation system have the same rate of change of the network loads, roads do not. In fact, it is reasonable to assume that a network manager has a coarser observation scale of a transportation system than the users, whose paths rapidly fluctuate. In practice,  this means that modifications in the network infrastructure occur on a much larger time scale than that of daily passengers' fluctuations.

Remarkably, we find that the Fourier decomposition of the loads yields a sufficient condition to determine whether the resulting optimal networks will contain loops or be a tree. Performing a numerical validation of our dynamics on synthetic networks we are also able to provide an analytical expression for the long-run conductivities.  Precisely, we find that the conductivities start oscillating around constant values at large time scales, and at certain frequencies that can be expressed in terms of those of the input loads. Furthermore, we define a Lyapunov functional for our dynamical formulation, allowing us to interpret stationary topologies as optimal networks, i.e., structures minimizing the global cost to build the graph. Finally, we examine a case study with loads that are the sum of decoupled harmonic oscillators, finding that the condition on the Fourier coefficients can be equivalently reformulated in terms of the loads' amplitudes and phases. We numerically investigate this last setup on the Bordeaux bus network.

\section{Time-varying loads in routing optimization on networks}
\label{sec:dyn}

Consider a network $\Graph$ with nodes $\Inode \in \Nodes$ and edges $\Iedge \in \Edges$, each of length $\ell_\Iedge > 0$. The orientation of the edges is conventionally assigned by the signed incidence matrix of the graph, with entries $\Inc_{\Inode \Iedge} = \pm 1$ if node $\Inode$ is the tail or the head of edge $\Iedge$, and $\Inc_{\Inode \Iedge} = 0$ otherwise. We consider a routing optimization problem on $\Graph$ setting time-varying mass loads $\NetMass(t) = \{ \NetMass_v(t) \}$ on nodes being the amount of mass either injected in ($\NetMass_v(t) >0$) or extracted from ($\NetMass_v(t) <0$) node $v$. Concretely, one could think of $S(t)$ as a time-dependent origin-destination vector of passengers moving in a transportation network, where mass entries correspond to the fraction of passenger flowing though stations. This allows us to write \kl~conservation law as
\begin{align}
\label{eqn:fast_kirch_law}
\sum_{u} L_{vu} (\Cond) \, \Press_u (t) = \NetMass_\Inode (t) \quad \forall \Inode \in \Nodes,  \forall t\geq 0\,,
\end{align}
where $\Cond = \{\Cond_\Iedge\}$ are the non-negative edge conductivities, $p(t) = \{ \Press_\Inode(t) \}$ are pressure potentials on nodes and $L_{vu}(\Cond) := \sum_{e} B_{ve} ({\Cond_\Iedge}/{\ell_\Iedge}) B_{ue}$ are the entries of the weighted Laplacian of the network \cite{klein1993resistance}. The conductivities can be interpreted as the capacities that the edges must have to allocate the mass loads acting on the nodes; thus we can consider them proportional to the edges' sizes. When considering passengers moving along a transportation network, $\mu$ can be seen as the width of a road or more generally a measure of the infrastructure's resources used to carry traffic flows.


\begin{figure}[t]
	\centering
	\includegraphics[width=\columnwidth]{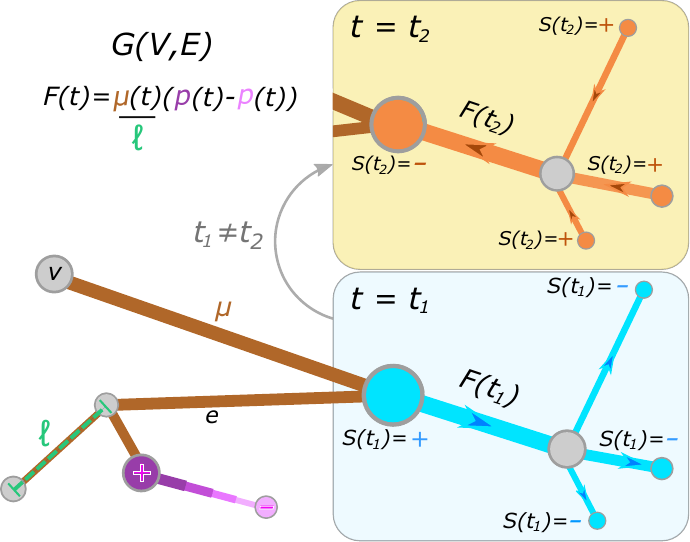}
	\caption{\label{fig:sketch} Schematic visualization of the problem. We highlight conductivities (brown),  the length of an edge (green), and the difference of pressure (purple) along an edge that triggers their fluxes. In the rightmost blue and yellow panels we depict two scenarios where the time-dependent loads $S(t)$ generate fluxes that move from the central part of the network to its periphery ($t=t_1$) and vice versa ($t=t_2 \neq t_1$). Node and edge widths are proportional to $S(t)$ and $F(t)$, respectively.}
\end{figure}


We propose a model in which the forcings $\NetMass(t)$ dictate the time evolution of the conductivities by means of a feedback dynamics. In particular, we couple \Cref{eqn:fast_kirch_law} with the system of with the system of ordinary differential equations (ODEs)
\begin{alignat}{2}
\label{eqn:fast_dyn}
\f{d \Cond_e (t)}{dt} &= \frac{ \Flux_e^2 (t) }{\Cond_e^{\gamma} (t)} - \Cond_e (t) \quad &&\forall \Iedge \in \Edges \\
\label{eqn:fast_dyn_init_cond}
\Cond_e  (0) &= m_\Iedge &&\forall \Iedge \in \Edges \,,
\end{alignat}
with $m_\Iedge > 0$ initial values. For a solution trajectory $\Cond(t)$, we define the fluxes $\Flux_\Iedge (t) \equiv F_e (\Cond(t),\NetMass(t) ) := \Cond_e(t) [\Press_\Inodethree (t) - \Press_\Inode (t)] /\ell_e$ for $\Iedge = (\Inodethree,\Inode)$, with $\Press_\Inode (t) \equiv  \Press_\Inode (\Cond(t),\NetMass(t)) := \sum_\Inodethree L^\dagger_{\Inode \Inodethree}(\Cond(t)) \NetMass_\Inodethree (t) $ the solution of \Cref{eqn:fast_kirch_law},  where $L^\dagger$ denotes the Laplacian pseudoinverse.  We assume that the system is isolated,  namely, $\sum_\Inode \NetMass_\Inode(t) = 0, \forall t \geq 0$, so $\Press(t)$ is a well-defined potential (see \cite{klein1993resistance}, Lemma 0). Specifically,  \citet{klein1993resistance} showed that $L$ generally is not invertible, but \Cref{eqn:fast_kirch_law} can be solved by the pseudoinverse within the subspace orthogonal to the unitary vector, that is, when $\sum_v S_v (t) = 0$.

In \Cref{eqn:fast_dyn}, the growth in time of the conductivities is proportional to the flux forcing term $F_e^2(t)$ with $\mu$ decaying exponentially when no flux flows though an edge. In practice, this corresponds to enlarging a road when many passengers travel along it, and reducing it when there is no traffic.  We illustrate this intuition with a schematic representation in \Cref{fig:sketch}.

The free parameter $0 < \gamma < 2$ tunes between different transportation mechanisms \cite{facca2020branching, lonardi2021designing,bonifaci2012physarum}. The case $\gamma < 1$ encourages mass consolidation on a few edges, $\gamma =1$ is shortest-path-like,  and  $1<\gamma < 2$ penalizes traffic congestion. 

Our dynamical formulation assumes continuous variables for fluxes and conductivities, but the mass $S(t)$ could be arbitrarily continuous or discrete. While this is valid in many scenarios (e.g., when modeling a large number of individuals), it may be limiting in cases where a discrete (or atomic) representation is necessary to capture fine-grain differences in the number of passengers. For this, one should consider alternative formulations and approaches, for instance, using message passing or belief propagation as in \cite{yeung2012competition, yeung2013networking, yeung2013physics, altarelli2015edge, de2014shortest, xu2021scalable}.

Finally, we remark that \Cref{eqn:fast_dyn} can be made scale independent with respect to the model variables by an opportune nondimensionalization that we describe in detail in \Cref{apx_sec:appendix_-1}.

\section{Model construction}
\label{sec:model_contsr}

\subsection{Slow adaptation of conductivities}
\label{ssec:slow_adapt_cond}

In several biological systems the adaptation time of organisms is much slower (weeks) than the characteristic time of the mass injected in the system (seconds) \cite{folkow1963studies,  granger1982role, widmer2007application, hu2012biological}.  In order to describe these organisms, a common approach is that of approximating the fast time-varying input loads with combinations of open and closed switch-like nodes with constant inflows, and to assume that the conductivities are regulated by an ensemble average of the pressures over different states of the loads \cite{hu2012biological,hu2013adaptation, katifori2010damage, corson2010fluctuations}. 

\begin{figure}[t]
	\centering
	\includegraphics[width=\columnwidth]{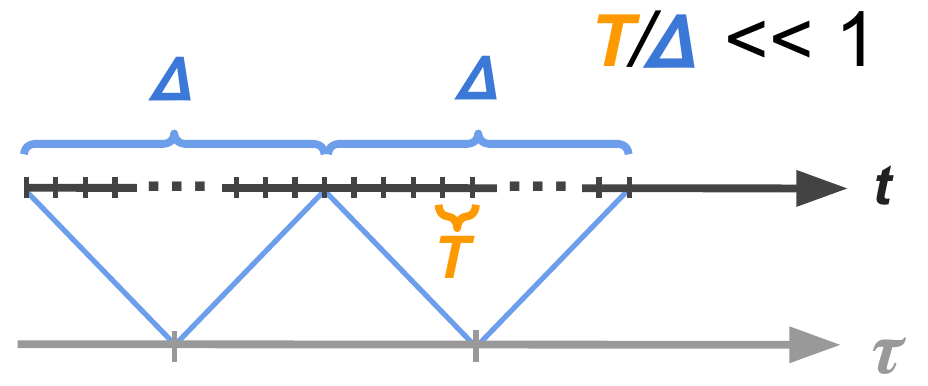}
	\caption{\label{fig:time_scales_scheme} Schematic representation of the different time variables.  The two arrows denote time scales $t$ and $\tau$. The time windows $\Delta$, large with respect to $t$, are denoted with blue curved brackets and the fast period $T$ in orange. Each window $\Delta$ along which we integrate the dynamics \Crefrange{eqn:fast_kirch_law}{eqn:fast_dyn_init_cond} contains a large number of periods $T$.}
\end{figure}

Instead, here we want to model the evolution of these slow adapting conductivities taking in account the time dependence of the loads.  We formalize this hypothesis by assuming: (i) the existence of a slow time scale $\tau$, with $\tau= K t$ and $K \gg 1$, and that (ii) in a fixed time window $\Delta$,  small with respect to the slow variable $\tau$ but large with respect to the time $t$, 
\begin{align}
\label{eqn:time_hom}
	\hat{\Cond}_\Iedge (\tau + t') \approx \hat{\Cond}_\Iedge (\tau) \quad \forall \Iedge \in \Edges, \forall t' \in [0, \Delta),\forall \tau \geq 0
\end{align}
holds for some conductivities $\hat{\mu} = \{ \hat{\mu}_e \}$ with the natural time of evolution being $\tau$.  Time scales are depicted in~\Cref{fig:time_scales_scheme}. We can interpret $t$ as seconds, $\Delta$ as days, and $\tau$ as months.
Such distinction between different natural time scales is observed in the interplay of rivers and tide loads in coastal delta formation \cite{konkol2021interplay}, where the assumption is that tides cycle much faster than the river channel adaptation, a distinction analogous to that between $t$ and $\tau$.

Finally, we assume that (iii) the evolution in $\tau$ of $\hat{\mu}$ is determined by the time integral average of the product of the mass loads.  Assumptions (i), (ii) and (iii) together lead to the definition of
\begin{equation}
\begin{split} \label{eqn:rhs_slow_def}
\hat{\Phi}_\Iedge(\hat{\Cond}, \tau) := \f{\hat{\Cond}_\Iedge^{2-\gamma}}{\ell^2_\Iedge} \sum_{\Inodethree \Inode} & A_{\Iedge \Inodethree}(\hat{\Cond}) A_{\Iedge \Inode}(\hat{\Cond}) \, \times \\ &\times \f{1}{\Delta} \int_{\tau}^{\tau + \Delta} \NetMass_\Inodethree (t) \NetMass_\Inode (t)\,dt  - \hat{\Cond}_\Iedge
\end{split}
\end{equation}
for all $\Iedge \in \Edges$ and $\tau \geq 0$, where we introduced $A_{\Iedge \Inode}(\hat{\Cond}) := \sum_{\Inodethree} \Inc_{\Iedge \Inodethree} L_{\Inode \Inodethree}^\dagger(\hat{\Cond}), \forall \Iedge \in \Edges,\forall \Inode \in \Nodes$.  The functional $\hat{\Phi}$ is the natural approximation of the right-hand side of \Cref{eqn:fast_dyn}, as shown in \Cref{apx_sec:appendix_0}. 


We define then a trajectory $\hat{\Cond}(\tau)$ as a solution of the dynamics
\begin{alignat}{2}
\label{eqn:slow_dyn}
\f{d \hat{\Cond}_\Iedge (\tau)}{d \tau} &= \hat{\Phi}_\Iedge ( \hat{\Cond} ( \tau ), \tau ) \quad &&\forall \Iedge \in \Edges\\
\label{eqn:slow_dyn_init_cond}
\hat{\Cond}_\Iedge (0) &= \hat{m}_\Iedge &&\forall \Iedge \in \Edges
\end{alignat}
with $\hat{m}_e > 0$ initial conditions. 

In general, $\hat{\Phi}$ is difficult to manipulate as the loads $\NetMass(t)$ may assume any arbitrary expression, possibly preventing the exact computation of the time integrals. For this reason, we investigate its behavior for a particular class of functions $\NetMass (t)$ that allows for analytical tractability.

\subsection{Periodicity of the loads}
We consider periodic loads  $\NetMass(t)$, with period $T$ small with respect to the fixed integration window $\Delta$ introduced in~\Cref{ssec:slow_adapt_cond}:
\begin{align}
\label{eqn:hyp_periodicity}
\NetMass_\Inode(t + T) = \NetMass_\Inode(t), \; T / \Delta \ll 1 \quad \forall \Inode \in \Nodes, \forall t \geq 0 \,.
\end{align}
This allows us to express each $\NetMass_\Inode(t)$ using its Fourier series $\NetMass_{\Inode}(t)=\sum_{n_{\Inode} \in \mathbb{Z}} c^{n_{\Inode}}_{\Inode} \exp({ \mathrm{i} \omega n_{v} t})$, with $\omega = 2\pi/T$. Substituting this into \Cref{eqn:rhs_slow_def} yields the pivotal result:
\begin{align}
\label{eqn:result_1_equation}
\f{1}{\Delta}\int_{\tau}^{\tau + \Delta} \NetMass_\Inodethree (t) \NetMass_\Inode (t)\,dt
 = C_{\Inodethree \Inode} + {O}\left( {\Delta} \right) \,,
\end{align}
holding for all $\Inodethree, \Inode \in \Nodes$ and $\tau \geq 0$. The matrix $C$ has entries $C_{\Inodethree \Inode} := \sum_{n_\Inode} {(c_\Inodethree^{n_\Inode})}^* c_\Inode^{n_\Inode}, \forall \Inodethree, \Inode \in \Nodes$, with $c^{*}$ denoting the complex conjugation of $c$. The term ${O}(\Delta)$ contains all negligible contributions $\varepsilon$, decaying  as $\varepsilon / \Delta \to 0$ for $\Delta \to + \infty$. For a detailed derivation of this result one can refer to \Cref{apx_sec:appendix_1}.

\subsection{Periodic-loads dynamics}

Combining \Cref{eqn:rhs_slow_def} and \Cref{eqn:result_1_equation}, we can build a dynamics for some new conductivities $\bar{\Cond} = \{\bar{\Cond}_e \}$, which evolve in the slow time scale $\tau$.  Precisely, we ignore negligible contributions in \Cref{eqn:result_1_equation} and define
\begin{alignat}{2}
\label{eqn:periodic_dyn}
\f{d \bar{\Cond}_\Iedge (\tau)}{d \tau} &= \bar{\Phi}_\Iedge ( \bar{\Cond}_\Iedge ( \tau ) ) \quad &&\forall e \in \Edges \\
\label{eqn:periodic_dyn_init_cond}
\bar{\Cond}_\Iedge (0) &= \bar{m}_\Iedge &&\forall e \in \Edges \,
\end{alignat}
with $ \bar{m}_\Iedge > 0$ initial conditions. The right hand side of \Cref{eqn:periodic_dyn} is such that $\hat{\Phi} \simeq \bar{\Phi}$ for $\Delta \gg 1$ and reads
\begin{align} \label{eqn:periodic_rhs_def}
\bar{\Phi}_{\Iedge}(\bar{\Cond}) := \f{\bar{\Cond}_\Iedge^{2-\gamma}}{\ell^2_\Iedge} \sum_{\Inodethree \Inode} A_{e \Inodethree} (\bar{\Cond}) A_{e \Inode} (\bar{\Cond}) C_{\Inodethree \Inode}  - \bar{\Cond}_\Iedge \quad \forall \Iedge \in \Edges \,.
\end{align}

An important point is that the problem in \Crefrange{eqn:periodic_dyn}{eqn:periodic_dyn_init_cond} is \textit{not} equivalent to the dynamics \Crefrange{eqn:fast_kirch_law}{eqn:fast_dyn_init_cond} with each $\NetMass_\Inode(t)$ integrated over $T$.  The latter case would imply that $C_{\Inode \Inodethree}$ had the form $C_{\Inodethree\Inode} = \tilde{\NetMass}_\Inodethree \tilde{\NetMass}_\Inode$, where $\tilde{\NetMass}_\Inode$ is the integral of $\NetMass_\Inode(t)$ over the period. This is only a particular case of the dynamics in \Crefrange{eqn:periodic_dyn}{eqn:periodic_dyn_init_cond}. 

Noticeably, in this case the condition $\text{rank} ( C ) = 1$ holds, i.e., since $C$ is symmetric, there exists a vector $y \in \mathbb{R}^{|\Nodes|}$ such that $C_{\Inodethree \Inode} = y_\Inodethree \, y_\Inode, \forall \Inodethree, \Inode \in \Nodes$. This is a sufficient condition for \Crefrange{eqn:periodic_dyn}{eqn:periodic_dyn_init_cond} to return a loopless network at convergence (see \Cref{apx_sec:appendix_2} for a proof), and confirms previous results observed for constant loads \cite{maritan,bohn2007structure,kirkegaard2020optimal}. 

However, this condition does not hold generally for any arbitrary choice of the loads, as $C$ may have a more general expression, in particular $ \text{rank} ( C ) > 1$. Moreover, the case of constant loads is not the only one where $ \text{rank} ( C ) = 1$. We provide an example of this  in \Cref{ssec:bordeaux}, where we explore the case of each $\NetMass_\Inode(t)$ being the sum of a finite number of harmonic oscillators.

\section{Characterization of the fast dynamics}
\label{sec:characterization}


\begin{figure*}[t]
	\centering
	\includegraphics[width=0.9\textwidth]{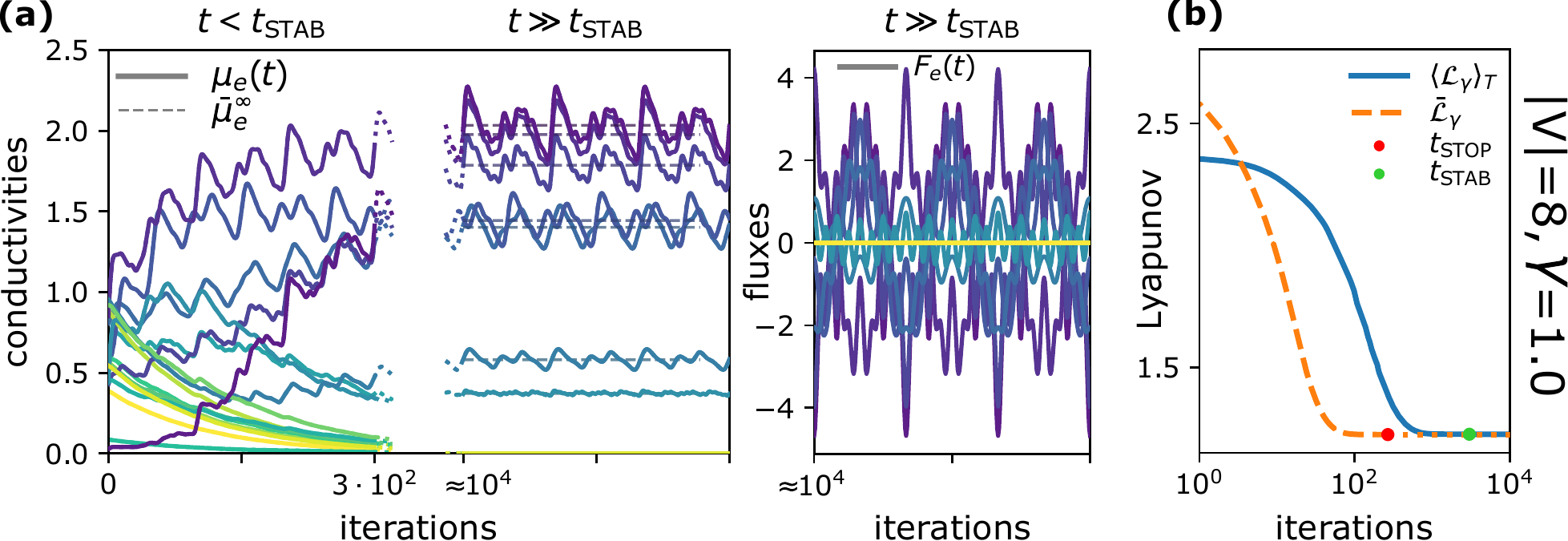}
	\caption{\label{fig:toymodel} Characterization of the fast conductivities $\Cond(t)$.  All results are computed on a synthetic network with $|V| = 8$ and setting $\gamma = 1.0$. (a) Fast conductivities $\Cond(t)$ and fluxes $\Flux(t)$ are drawn with solid lines, stationary solutions $\bar{\Cond}^\infty$ are dashed. Labels on the $x$ axis correspond to the number of iterations of the numerical discretization of \Crefrange{eqn:fast_kirch_law}{eqn:fast_dyn_init_cond}. Conductivities are depicted in two time windows, before and after their stabilization time $t_\text{STAB}$. Fluxes are drawn only for $t \gg t_\text{STAB}$.  Colors denote different edges. (b) Evolution of $\bar{\mathcal{L}}_\gamma$ and of $\langle {\mathcal{L}}_\gamma \rangle_T$ in time. The green and the red circles denote $t_\text{STAB}$ and $t_\text{STOP}$, respectively. 
}
\end{figure*}


Finding an analytical expression for the fast conductivities $\Cond(t)$ solutions of \Crefrange{eqn:fast_kirch_law}{eqn:fast_dyn_init_cond} cannot be done by directly solving the dynamics, because of the non linear dependence on $\Cond(t)$ in the Laplacian pseudoinverse. Nevertheless, here we propose an argument to characterize their long-time behavior.

We support our findings with an empirical validation on synthetic networks built taking the Delaunay triangulation of $|V|$ nodes placed at random in the unit square. In our experiments, we set  $|V| = 2^i$, with $i = 3,\dots, 9$. The vector of loads $S(t)$ is $S(t) := 20\, S_1(t) + 10\,  S_4(t) + 5\, S_8(t)$, where each factor is defined as $S_n(t) := q_n \cos (\omega n t)$, with amplitudes extracted at random from a $|V|$-dimensional Dirichlet distribution as $q_n \sim \mathcal{D}(\alpha = 1) - 1/|V|$ [so that $\sum_\Inode \NetMass_\Inode (t) = 0$ $\forall t \geq 0$] and $n = 1,4,8$. The period has been conventionally set to have $\omega = 2 \pi$.

We observe that the evolution of the fast conductivities is typically divided in two phases, as shown in \Cref{fig:toymodel}a. First, the conductivities undergo a stabilization transient for $t < t_\text{STAB}$,  where they strongly depend on their initial conditions $m_\Iedge$ and significantly change their mean values. Then, when $t > t_\text{STAB}$, the conductivities reach a plateau and oscillate around fixed values. More precisely, either they move around mean values that are far from zero and preserve their oscillatory nature for all times, or they decay to zero with negligible oscillations that are progressively damped as $t$ increases. These experimental observations suggest the following ansatz for the stabilized solutions, for all $t > t_\text{STAB}$ and $\Iedge \in \Edges$:
\begin{align}
\label{eqn:ansatz_fast}
\Cond_\Iedge(t) = a_\Iedge + b_\Iedge(t) \; \text{s.t.} \; a_e =  \text{const.}, \; b_e(t+T)=b_e(t) \,.
\end{align}

We compare solutions of the new dynamics \Crefrange{eqn:periodic_dyn}{eqn:periodic_dyn_init_cond} with those of \Crefrange{eqn:fast_kirch_law}{eqn:fast_dyn_init_cond} [see \Cref{fig:toymodel}a for an example]. In the figure the conductivities $\Cond(t)$ are oscillating around the constant values of $\bar{\Cond}(\tau)$ reached at convergence, which we denote with $\bar{\Cond}^\infty = \{ \bar{\mu}_e^\infty \}$.  Motivated by this empirical observation,  we set
\begin{align}
a_e = \bar{\Cond}_\Iedge^\infty \quad\forall \Iedge \in \Edges \,.
\end{align}

We experimentally notice that also the fluxes start to oscillate around a constant value after a first stabilization time interval [see \Cref{fig:toymodel}a]. We use this evidence to deduce (see \Cref{apx_sec:appendix_3}) that the main oscillatory modes of the conductivities are resonant with the squared fluxes, and have the form
\begin{align}
\label{eqn:epression_b}
b_\Iedge(t) = \sum_{n,m \in \mathcal{N}} b_\Iedge^m b_\Iedge^n \exp ( \mathrm{i} \omega (n+m) t) \quad \forall \Iedge \in \Edges \,,
\end{align}
with $\mathcal{N} := \{n_\Inode \}$ set of the Fourier modes of the loads.  Hence, the conductivities oscillate with modes determined by those of the loads. This result is supported by several numerical experiments (see \Cref{apx_sec:appendix_3} for details).

Remarkably,  these numerical experiments also serve as a validation for hypothesis (ii) in \Cref{ssec:slow_adapt_cond}. In fact, for any sufficiently slow time $\tau$, the conductivities fluctuate around a constant value, suggesting the possibility of neglecting their fast oscillatory nature when studying asymptotics of \Crefrange{eqn:fast_kirch_law}{eqn:fast_dyn_init_cond}.

\subsection{Candidate Lyapunov functional}

We empirically observe [see \Cref{fig:toymodel}b] that our new dynamics \Crefrange{eqn:periodic_dyn}{eqn:periodic_dyn_init_cond} admits a candidate Lyapunov functional reading
\begin{align}
 \label{eqn:functional}
\bar{\mathcal{L}}_\gamma (\bar{\mu}) := \frac{1}{2} \sum_\Iedge \frac{\ell_\Iedge}{\bar{\mu}_\Iedge} \bar{F}^2_e(\bar{\mu}) + \frac{1}{2\gamma} \sum_\Iedge \ell_\Iedge \bar{\mu}_\Iedge^\gamma \,,
\end{align}
where for each edge $e$ we define the squared slow fluxes $\bar{F}^2_\Iedge := (\bar{\Cond}_\Iedge^2 / \ell_\Iedge^2 ) \sum_{\Inodethree \Inode } A_{\Iedge \Inodethree }(\bar{\Cond}) A_{\Iedge \Inode }(\bar{\Cond}) C_{\Inodethree \Inode}$.

Noticeably, if $\text{rank}(C) = 1$ holds, it is possible to formally prove that $\bar{\mathcal{L}}_\gamma (\bar{\mu})$ is a well-defined Lyapunov functional for \Crefrange{eqn:periodic_dyn}{eqn:periodic_dyn_init_cond} (see \Cref{apx_sec:appendix_2} for detailed derivations). In addition,  we can interpret the functional as in Lonardi \textit{et al.} \cite{lonardi2021designing} for multicommodity optimal transport. Namely, the Lyapunov is the sum of a dissipation cost, the first addend in \Cref{eqn:functional},  with an infrastructural cost, the price needed to build the transport network.

We notice empirically that the functional reaches a plateau at $t_\text{STOP}$, defined as the time for which $\Delta \bar{\mathcal{L}}_\gamma / \bar{\delta t} < \varepsilon$ is satisfied, with $\Delta \bar{\mathcal{L}}_\gamma := |(\bar{\mathcal{L}}_\gamma)^{\tau+1} + (\bar{\mathcal{L}}_\gamma)^{\tau} | / (\bar{\mathcal{L}}_\gamma)^{\tau+1} $, where the upper indices are consecutive iterations in the finite-difference discretization of \Crefrange{eqn:periodic_dyn}{eqn:periodic_dyn_init_cond}. In all our experiments, we set $\bar{\delta t} = 0.1$ as time step of a Forward Euler method, and the convergence threshold to $\varepsilon = 10^{-5}$.

Additionally, we observe that the candidate Lyapunov functional $\bar{\mathcal{L}}_\gamma$ converges to a value that is the same achieved by the running average functional over the period $T$:
\begin{align}
\label{eqn:functional_running}
\langle \mathcal{L}_\gamma \rangle_T := \f{1}{T}\int_t^{t+T} \bup{  \frac{1}{2} \sum_\Inode p_v(\Cond) \NetMass(t') + \frac{1}{2\gamma} \sum_\Iedge \ell_\Iedge \Cond_\Iedge^\gamma } dt'
\end{align}
with $\Cond$ that is evaluated along solution trajectories of \Crefrange{eqn:fast_kirch_law}{eqn:fast_dyn_init_cond}. The functional \Cref{eqn:functional_running} reaches a plateau at the stabilization time $t_\text{STAB}$, when the fast conductivities $\Cond(t)$ start oscillating around constant values. Remarkably, in \Cref{fig:toymodel}b we see that $t_\text{STAB} \gg t_\text{STOP}$, which is due to the fact that the time step $\delta t$ for the numerical discretization of \Crefrange{eqn:fast_kirch_law}{eqn:fast_dyn_init_cond} has to be set much lower than $\bar{\delta t}$ in order to capture the oscillatory nature of the loads. In our experiments we set it to $\delta t = \bar{\delta t}/10$. A practical consequence of this is that the discretization of \Crefrange{eqn:periodic_dyn}{eqn:periodic_dyn_init_cond} is a fast and scalable alternative to extract the conductivities around which long-run solutions of \Crefrange{eqn:fast_kirch_law}{eqn:fast_dyn_init_cond} stabilize.

Because of this analogy between an optimal transport (functional minimization) setup and the solutions of our dynamical system, we can interpret the networks determined from the dynamics in \Crefrange{eqn:periodic_dyn}{eqn:periodic_dyn_init_cond} as optimal topologies  minimizing the infrastructural and dissipation cost. These networks can also be obtained by averaging long-run solutions of the original dynamics in \Crefrange{eqn:fast_kirch_law}{eqn:fast_dyn_init_cond}. In fact, as discussed in \Cref{sec:characterization}, long-run trajectories of \Crefrange{eqn:fast_kirch_law}{eqn:fast_dyn_init_cond} oscillate around asymptotics of the newly defined dynamical system in \Crefrange{eqn:periodic_dyn}{eqn:periodic_dyn_init_cond}.

\section{Generation of loops}
\label{sec:example}

\subsection{Conditions for the generation of loops in closed form}
\label{ssec:conditions_closed_form}

If $C$ has $\text{rank}(C) = 1$, i.e., $C_{\Inode \Inodethree} = y_\Inodethree \, y_\Inode$ for some $y \in \mathbb{R}^{|\Nodes|}$, the dynamics \Crefrange{eqn:periodic_dyn}{eqn:periodic_dyn_init_cond} produces trees at convergence. One  trivial case where this holds is when the loads $\NetMass(t)$ are static, i.e., constant for all times. However, this is not the only setting where $\text{rank}(C) = 1$ is satisfied. In particular, there are cases where such a condition holds but $\NetMass$ do change in time. 

Here, we explore a case of study proposing an ansatz where the loads are the sum of decoupled harmonic oscillators: 
\begin{align}
\label{eqn:ansatz_finite_cos_sum}
\NetMass_\Inode (t) =  \sum_{i = 1}^{N_\Inode} A^i_\Inode \cos  ({\omega} \, n^i_\Inode \, t + \phi^i_\Inode ) + d_\Inode \quad \forall \Inode \in \Nodes\,,
\end{align}
with $\omega = 2\pi / T$, $n^i_\Inode,  N_\Inode \in \mathbb{N}$, and $A_\Inode^i, d_\Inode \in \mathbb{R}$.  By construction, these loads are periodic in $T$, hence we compare them with their Fourier series representation $
\NetMass_\Inode(t) = {a^0_\Inode/2}+\sum_{n_\Inode \geq 1} a_\Inode^{n_\Inode} \cos \left( \omega n_\Inode t+\varphi^{n_\Inode}_\Inode \right)$.  Equating this expression with \Cref{eqn:ansatz_finite_cos_sum} yields
\begin{align} \label{eqn:importan_rel_harmonic}
c_\Inode^{n_\Inode} = \f{A_\Inode^{i}}{2} \exp ( {\mathrm{i} \phi_\Inode^{i}} ) \, \delta_{n_\Inode n_\Inode^i} \quad \forall \, n_\Inode \in \mathbb{N} \,,
\end{align}
where we conventionally set $\phi_\Inode^0 = 0, \forall \Inode \in \Nodes$ and where only a finite number of Fourier coefficients are different from zero, given that the sum in \Cref{eqn:ansatz_finite_cos_sum} is finite. 

The goal here is to express $\text{rank}(C) = 1$ in terms of $\{ A_\Inode^i \}$, $\{ n_\Inode^i \}$, and $\{ \phi^i_v \}$, amplitudes, modes and phases of the harmonic oscillators, respectively.  To do that, we start by noting that  $\text{rank}(C) = 1$ is satisfied if and only if $C_{\Inodethree \Inode} = y_\Inodethree y_\Inode$, $\forall \Inodethree, \Inode \in \Nodes$, with $y_v = \pm \sqrt{C_{\Inode \Inode}}$, and where the plus or minus signs have to be determined among $2^{\Nnode}$ possible choices in such a way that $\sum_\Inode y_\Inode = 0$ (see \Cref{apx_sec:appendix_4}).

Defining the complex vectors $\nu_\Inode = \{c^{n_v}_v\}$ with entries of the Fourier coefficients in \Cref{eqn:importan_rel_harmonic}, we rewrite $C_{\Inodethree \Inode}  =\pm \sqrt{C_{\Inodethree \Inodethree}} \sqrt{C_{\Inode \Inode}} $ as $\nu_\Inodethree \cdot \nu_\Inode = \pm ||\nu_\Inodethree|| ||\nu_\Inode || $, where the centered dot denotes the complex dot product and $|| \cdot ||$ is its correspondent norm.  Thus, the rank condition on $C$ can be reformulated in terms of an equivalent linear dependence condition of the form $\nu_\Inode = \lambda \nu_\Inodethree$ between the vectors $\nu_v$, $\Inode \in \Nodes$, and for $\lambda \neq 0$. Finally, substituting \Cref{eqn:importan_rel_harmonic} in this  linear dependence condition leads to the following main result. 

\begin{propo}
Let the time-dependent loads $S(t)$ injected in the network nodes be as in \Cref{eqn:ansatz_finite_cos_sum}. If the following hold, then, for any $\gamma \leq 1$,  a stationary solution of \Crefrange{eqn:periodic_dyn}{eqn:periodic_dyn_init_cond} is a tree:
\begin{enumerate}
\item $\phi_\Inode^i = \phi_\Inodethree^i + k\pi, \, k \in \mathbb{Z}$, i.e. sources and sinks are in phase,
\item $A_v^i \delta_{n_\Inode n_\Inode^i} = \lambda (-1)^k A_u^i \delta_{n_\Inodethree n_\Inodethree^i}$ (implying that $N_\Inode = N$ for all $\Inode$),
\end{enumerate}
\end{propo} 

For a formal justification of this result see \Cref{apx_sec:appendix_4}.

\subsection{Numerical tests on the Bordeaux bus}
\label{ssec:bordeaux}


\begin{figure*}[t]
 \centering
 \includegraphics[width=0.9\textwidth]{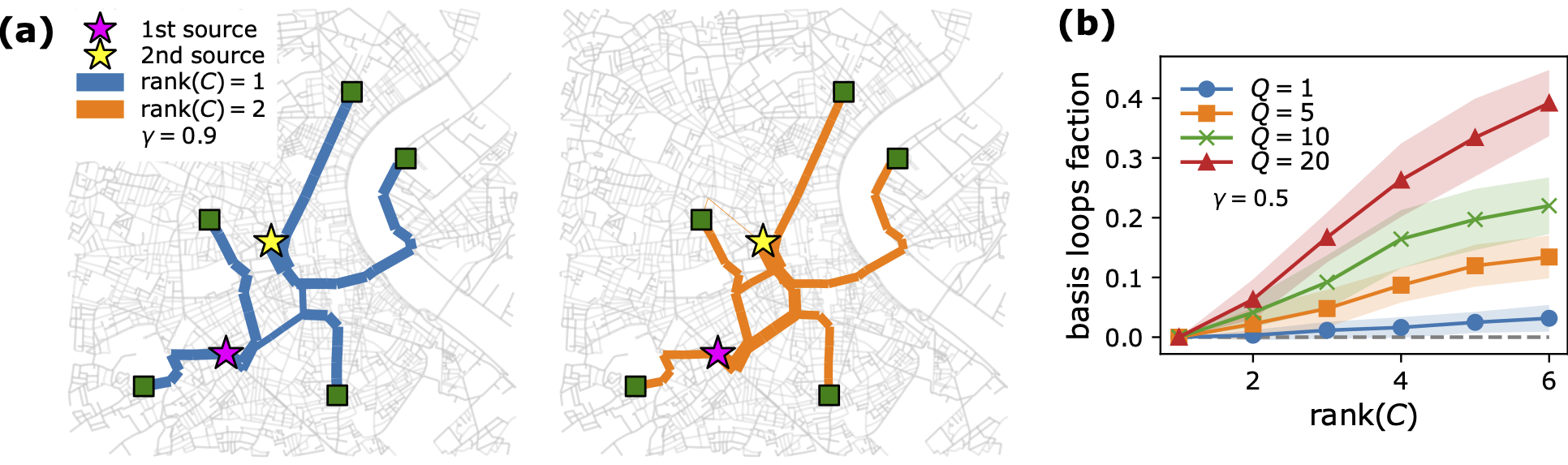}
 \caption{\label{fig:realnetwork} Bordeaux bus optimal transport network. (a) Network visualization. Input loads have been built as described in \Cref{ssec:bordeaux}. The tree network originated by $C$ with rank $1$ is plotted in blue, the loopy topology in orange. The yellow and the magenta stars denote the geographical location of the two loads, the green squares those of the sinks. Here the width of edges corresponds to slow conductivities at convergence $\bar{\Cond}_\Iedge^{\infty}$. Results are plotted for $\gamma = 0.9$ (b) Basis loop fraction against $\text{rank}(C)$. Points correspond to averages over $100$ runs of the experiments where positions of the sources and sinks are extracted at random. Shaded regions denote their standard deviations. Results are displayed for $\gamma = 0.5$.}
\end{figure*}


In order to test the rank condition on $C$ we design two experiments on the real network of the buses of Bordeaux. The network topology has been constructed focusing on a central region of the city, and using data collected from \cite{kujala2018collection}. Here we assume that the loads, representing passengers entering or exiting the network, vary much faster than the conductivities. These latter quantities can be thought of as the size of the roads that a network manager needs to design; thus we can safely assume their evolution to happen on a larger time scale with respect to that of $S(t)$. 

First, we design a simulation with two source nodes $v_1$ and $v_2$ [the stars in \Cref{fig:realnetwork}a], and five sinks (the green squares in \Cref{fig:realnetwork}a) that we extract at random among the nodes of the network. Then we consider two cases where the sources are built in such a way that (i) $\text{rank}(C)=1$ and (ii) $\text{rank}(C) = 2$. These are, respectively,
\begin{itemize}
\item[(i)] $S_{v_1}(t) = S_{v_2}(t) = 100 \cos(\omega t)$, with $\omega = 2 \pi$;
\item[(ii)] $S_{v_1}(t) = 100 \cos(\omega_1 t)$, with $\omega_1 = 2 \pi$\\$S_{v_2}(t) = 100 \cos(\omega_2 t)$, with $\omega_2 = 4 \pi$. 
\end{itemize}
All the sinks $\Inodethree \neq \Inode_1,\Inode_2$ have loads $S_u(t) = - [S_{v_1}(t) + S_{v_2}(t)]/5$ in both cases, to ensure conservation of mass.

We expect that in the first case the network extracted from \Crefrange{eqn:periodic_dyn}{eqn:periodic_dyn_init_cond} with $\gamma \leq 1$ is a tree. In the second case the network can possibly contain loops. We run the dynamics setting $\gamma = 0.9$ and we display our findings in \Cref{fig:realnetwork}a. The empirical results reflect our predictions: the blue network (the first case) is a tree. In contrast,  in the orange network (the second case) loops emerge.

We further validate our results on the bus network of Bordeaux with a second experiment. We assign the loads $S_v(t) = \sum_{i=1}^n S^i_v(t)$, with $S_v^i = (100 / |Q_n|) \cos(\omega i t)$,  to a set $Q_n$ of randomly extracted nodes, and $S_v^i = -[100 / (|V|-|Q_n|)] \cos(\omega i t)$ to the remaining ones. The modes are $ n = 1,\dots,6$, while the number of nodes which are randomly extracted for each $n$ are $Q:=|Q_n| = \{1,5,10,20\}$. We set again $\omega = 2\pi$. 

Exploiting the exact relation that the matrix $C$ has with the modes of the loads (see \Cref{ssec:conditions_closed_form}), it is possible to see that our particular construction of $\NetMass(t)$ gives ranks ranging in $1 \leq \text{rank}(C) \leq 6$.

We show our results in \Cref{fig:realnetwork}b, where we plot the fraction of basis loops of the network against the rank of $C$. The dynamics is executed for $\gamma = 0.5$ and the random extraction of the forcings has been varied over $100$ runs. In the plot, it is clearly visible that for all values of $Q$, the fraction of basis loops is zero at $\text{rank}(C) = 1$. Moreover, we can see that when we increase the complexity of the problem, i.e., when $\text{rank}(C)$ grows,  the values attained on the $y$ axis also increase. This suggests that the rank of the $C$ can be used as a qualitative proxy to predict the number of loops in the optimal transport network. Finally, as one could intuitively expect, the basis loop fraction increases with $Q$, i.e., with the number of nodes where mass is injected or extracted.

\section{Conclusions and Outlooks}

Routing models on networks are relevant to study many real-world problems. While most of the works in the current literature consider stationary setups, i.e., the inflows injected in the network do not change in time \cite{hu2013adaptation, ronellenfitsch2019phenotypes, lonardi2021designing, adinoyi2021optimal, bonifaci2012physarum, bonifaci2013short, facca2016towards, facca2019numerics, facca2020branching, facca2020physarum}, few recent works investigate time-varying loads and the majority of these models study solely the averaged evolution of the networks' variables \cite{hu2013adaptation, katifori2010damage, corson2010fluctuations, konkol2021interplay}.

In this work we analyzed a dynamical system where the conductivities are regulated by time-varying mass inflows.  Motivated by empirical observations \cite{folkow1963studies,  granger1982role, widmer2007application, hu2012biological}, we assumed the existence of auxiliary conductivities that have response times which are much slower than those of the loads. Furthermore, in order to make the problem analytically tractable, we supposed that all the loads injected in nodes are periodic,  in a period that is substantially smaller than the adaptation time of the new conductivities. These two hypothesis together allowed us to deduce a dynamics where the evolution of the systems is solely regulated by an input matrix constructed using the Fourier series expansion of the loads.

The resulting dynamics allowed us to derive the main findings of our work.  In detail, combining theoretical arguments with empirical evidence on synthetic networks, we found an expression for the long-run solutions of the original dynamics, which cannot otherwise be obtained by simply solving the original dynamics. These long-run solutions are the sum of stationary components, equal to the asymptotics of the dynamics we constructed, and an oscillatory one. This second contribution can be expressed as the sum of periodic signals, with modes related to those of the loads.
Moreover, we discussed a sufficient condition on the loads that determines when optimal transport networks can be loopless. Such a condition was numerically validated on the Bordeaux bus network. Finally, our dynamics can be connected to an optimization setup, as shown by the proposed candidate Lyapunov functional. As a result, asymptotic trajectories of our dynamics minimize the total cost needed to build the network infrastructure.

Importantly, the numerical discretization of the dynamics we proposed in this work can be used as an efficient method to rapidly converge to average long-run solutions of the original dynamics.

Our results can be extended in several ways. For instance, it would be interesting to investigate different types of input loads that relax the periodicity hypothesis and use this to analyze the behavior of the conductivities in different problems' settings. Similarly, it would be interesting to explore how this formalism adapts to multilayer networks, where passengers can enter different stations corresponding to different transportation modes \cite{adinoyi2021optimal}. Another relevant application could be that of integrating our findings with the recent work of Baptista \textit{et al.} \cite{baptista2021convergence}, where the authors studied how topological properties of the transport network change in time, as we approach stationary configurations, and how these reflect on the shape of the conductivities.

While our work constitutes a step towards extending the formalism of capacitated networks to time-dependent loads, it is important to remark that our findings are valid in a particular time limit. Specifically, this is the scenario where conductivities slowly evolve with the integral average of periodic forcings, as introduced in \Cref{sec:model_contsr}.  It is not clear how the theoretical analysis presented in this work could be adapted to scenarios where loads and conductivities evolve with the \textit{same} time scale. This could be an interesting avenue for future work.
Another interesting direction could be that of considering additional constraints on the evolution of the conductivities, which are not currently included in our model. For instance, one could introduce a threshold capacity above which the edge traffic saturates, causing blockage of roads.

Altogether,  we believe that our results enrich the current knowledge on network routing problems with time-varying input loads and have immediate practical implications. In fact our model is deterministic, since there is only one single realization of the inputs, and thus adequate to model real-world scenarios where time-dependent loads are measured quantities, e.g., the amount of passengers traveling in a metro (which can be easily tracked), without the need of stochastic formulations that require the introduction of probability distributions that are hard to characterize.

To facilitate practitioners in using our model, we have made the algorithmic implementation publicly available \cite{git_repo}.

\section{Acknowledgements}

The authors thank the International Max Planck Research School for Intelligent Systems (IMPRS-IS) for supporting Alessandro Lonardi.

\appendix

\section{Nondimensionalization of the model}
\label{apx_sec:appendix_-1}

Here we show how our model can be made dimensionless, i.e., constants can be removed by appropriately rescaling dimension-dependent quantities.  We start from the dimension-dependent ODEs:
\begin{align}
\label{eqn:dimensional_ode}
\frac{d\tilde{\mu}_e (\tilde{t}) }{d\tilde{t}} = a \frac{\tilde{F}^2_e (\tilde{t}) }{\tilde{\mu}^\gamma_e (\tilde{t}) } - b \tilde{\mu}_e(\tilde{t})  \quad \forall \Iedge \in \Edges \,,
\end{align}
with $a$ and $b$ coefficients with appropriate dimensions. We then choose the nondimensionalization
\begin{alignat}{2}
\label{eqn:dim}
t &:= \tilde{t} / t_c \\
\mu_e &:= \tilde{\mu}_e / \mu_c \qquad &&\forall e \in E \\
\label{eqn:dim1}
S_v(t) &:= \tilde{S}_v(\tilde{t}) / S_c &&\forall v \in \Nodes \,
\end{alignat}
where $S_c$ is the characteristic unit of $S$.  Substituting \Crefrange{eqn:dim}{eqn:dim1} in Kirchhoff's law yields $F_e(t) = \tilde{F}_e(\tilde{t})/S_c$, $\forall e \in E$, with $F(t)$ adimensional fluxes.

Recasting all adimensional variables in \Cref{eqn:dimensional_ode}, we get
\begin{align}
    \frac{d {\mu_e (t)} }{d {t}} = a \left( \frac{t_c S_c^2}{\mu_c^{\gamma + 1}} \right)\frac{F_e^2({t})}{\mu_e^\gamma(t)} - b t_c \, {\mu_e(t)} \quad \forall e \in E \,,
\end{align}
showing that, to recover \Cref{eqn:fast_dyn}, we can set
\begin{align}
t_c &= 1/b \\
\mu_c^{\gamma + 1}/S_c^2 &= a/b \,.
\end{align}
We note that a procedure for the nondimensionalization of a model similar to ours can be found in \cite{ronellenfitsch2019phenotypes} (Supp. Mat. Sec. II).

\section{Derivation of \Cref{eqn:rhs_slow_def}}
\label{apx_sec:appendix_0}

In order to define \Cref{eqn:rhs_slow_def}, we perform the calculations on the right hand side of \Cref{eqn:fast_dyn},
\begin{widetext}
\begin{alignat}{2}
\int_{\tau}^{\tau + \Delta} \Phi_\Iedge(\Cond(t),t) \, dt &:= \frac{1}{\Delta} \int_{\tau}^{\tau + \Delta} \Cond^{-\gamma}_\Iedge(t) \Flux^2 _\Iedge(t)- \Cond_\Iedge(t) \, dt \\
\label{eqn_apx:pass_1}
&= \frac{1}{\Delta}  \int_{\tau}^{\tau + \Delta} \bigg( \frac{\Cond^{2-\gamma}_\Iedge(t)}{\ell_\Iedge^2} \sum_{\Inodethree \Inode mn} \Inc_{m \Iedge} \Inc_{n \Iedge} \Lap^\dagger_{\Inodethree m}(\Cond(t)) \Lap^\dagger_{\Inode n}(\Cond(t))  S_\Inodethree(t) S_\Inode(t) - \Cond_\Iedge(t)  \bigg) \, dt \\
\begin{split}
&\overset{t' = t - \tau}{=} \frac{1}{\Delta}  \int_{0}^{\Delta}  \bigg( \frac{\Cond^{2-\gamma}_\Iedge(\tau + t')}{\ell_\Iedge^2} \sum_{\Inodethree \Inode mn} \Inc_{m \Iedge} \Inc_{n \Iedge} \Lap^\dagger_{\Inodethree m}(\Cond(\tau + t')) \Lap^\dagger_{\Inode n}(\Cond(\tau + t')) \, \times  \\  &\qquad\qquad\qquad\qquad\qquad\qquad\qquad\qquad\qquad\qquad\qquad  \times  S_\Inodethree(\tau + t') S_\Inode(\tau + t') - \Cond_\Iedge(\tau + t')  \bigg) \, dt'
\end{split}\\
\label{eqn_apx:pass_2}
& \overset{\textrm{(ii)}}{\approx} \frac{\Cond^{2-\gamma}_\Iedge(\tau)}{\ell_\Iedge^2} \sum_{\Inodethree \Inode} A_{\Iedge \Inodethree}(\Cond(\tau))A_{\Iedge \Inode}(\Cond(\tau)) \frac{1}{\Delta}  \int_0^\Delta S_\Inodethree(\tau + t') S_\Inode(\tau + t') - \Cond_\Iedge(\tau + t')  \, dt' \\
&\overset{t = t' + \tau}{=} \frac{\Cond^{2-\gamma}_\Iedge(\tau)}{\ell_\Iedge^2} \sum_{\Inodethree \Inode} A_{\Iedge \Inodethree}(\Cond(\tau))A_{\Iedge \Inode}(\Cond(\tau)) \frac{1}{\Delta}  \int_\tau^{\tau + \Delta} S_\Inodethree(t) S_\Inode(t) - \Cond_\Iedge(t)  \, dt \\
& =: \hat{\Phi}_\Iedge(\Cond(\tau),\tau) \,,
\end{alignat}
\end{widetext}
which are valid for all $\Iedge \in \Edges$. In detail, in \Cref{eqn_apx:pass_1} we used the definition of the fluxes $\Flux_\Iedge(t) := \Cond_\Iedge(t) [p_{\Inodethree}(t) - p_\Inode(t)] / \ell_\Iedge, \forall \Iedge \in \Edges$ and evaluated the pressure solving \kl~law, i.e. $p_\Inode(t) := \sum_\Inodethree L_{\Inode \Inodethree}^\dagger(\Cond(t)) S_\Inodethree(t), \forall \Inode \in \Nodes$. The second important step is in \Cref{eqn_apx:pass_2}, where we used hypothesis (ii) in \Cref{ssec:slow_adapt_cond}, namely the approximation in \Cref{eqn:time_hom}, to carry the conductivities out of the time integral, and we introduced $A_{\Iedge \Inode}(\Cond(t)) := \sum_{\Inodethree} \Inc_{\Iedge \Inodethree} \Lap^\dagger_{\Inode \Inodethree}(\Cond(t)), \forall \Iedge \in \Edges, \forall \Inode \in \Nodes$.

\section{Derivation of \Cref{eqn:result_1_equation}}
\label{apx_sec:appendix_1}

We enforce the hypothesis of periodicity of the loads, i.e. $\NetMass_\Inode(t) = \NetMass_\Inode(t + T)$,  with $T/\Delta \ll 1$, and we parametrize the integration window $\Delta$ as $\Delta = K T$, $K \gg 1$. This allows us to split the integral in \Cref{eqn:result_1_equation} into two separate contributions.  In detail, making the reasonable hypothesis that $\NetMass_\Inodethree (t) \NetMass_\Inode (t)$ is bounded by $M < +\infty$ for all $t \geq 0$ and for all $\Inodethree,\Inode \in \Nodes$, we can write
\begin{widetext}
\begin{align}
\label{eqn:quiv_periodic_forcing}
\f{1}{\Delta}\int_{\tau}^{\tau + \Delta} \NetMass_\Inodethree (t) \NetMass_\Inode (t)\,dt &=
\f{1}{\Delta} \int_{\tau}^{\tau +  \Delta} \sum_{n_\Inodethree n_\Inode} {(c_\Inodethree^{n_\Inodethree})}^* c_\Inode^{n_\Inode} \exp \left(\mathrm{i} \omega (n_\Inode - n_\Inodethree)t \right) \, dt \\
&= \sum_{n_\Inodethree n_\Inode} {(c_\Inodethree^{n_\Inodethree})}^* c_\Inode^{n_\Inode} \bigg( \sum_{k=1}^{\floor*{K}} \mathcal{I}_k(n_u,n_v) +  \mathcal{I}_K(n_u,n_v) \bigg) \,,\\
\mathcal{I}_k(n_u,n_v)  &:= \f{1}{\Delta} \int_{\tau + (k-1)T}^{\tau + k T}  \exp \left(\mathrm{i} \omega (n_\Inode - n_\Inodethree)t \right)\, dt \quad \forall k = 1,\dots,\floor*{K}\\
\mathcal{I}_K(n_u,n_v)  &:= \f{1}{\Delta}  \int_{\tau + \floor*{K} T}^{\tau + K T}  \exp \left(\mathrm{i} \omega (n_\Inode - n_\Inodethree)t \right) \, dt \,.
\end{align}
\end{widetext}

Hence, we separate the first $\floor*{K}$ integrals over the period $T$ from the last one in $(\floor*{K}T,  KT)$. Since  $K \gg 1$, the first $\floor*{K}$ contributions can be evaluated as
\begin{widetext}
\begin{align}
\sum_{n_\Inodethree n_\Inode} {(c_\Inodethree^{n_\Inodethree})}^* c_\Inode^{n_\Inode} \sum_{k=1}^{\floor*{K}} \mathcal{I}_k(n_u,n_v) &= \f{\floor*{K}}{K} \sum_{n_\Inodethree n_\Inode} {(c_\Inodethree^{n_\Inodethree})}^* c_\Inode^{n_\Inode} \delta_{n_\Inodethree  n_\Inode} \\ &= \sum_{n_\Inode} {(c_\Inodethree^{n_\Inode})}^* c_\Inode^{n_\Inode} + {O}\left( \Delta \right) \,,
\end{align}
\end{widetext}
with $\delta_{ij}$ being the Kronecker delta for two indices $i$ and $j$. As for the second term, in the limit $K \gg 1$ we can write
\begin{align}
\left| \sum_{n_\Inodethree n_\Inode} {(c_\Inodethree^{n_\Inodethree})}^* c_\Inode^{n_\Inode} \mathcal{I}_K(n_u,n_v)  \right| \leq \f{K - \floor*{K}}{K} M \sim {O}\left( {\Delta}\right) \,,
\end{align}
showing that integrals over the small interval $(\floor*{K}T, KT)$ are negligible for a large integration window.

\section{Sufficient condition on the rank for optimal trees}
\label{apx_sec:appendix_2}

We discuss in detail the sufficient condition $\text{rank}(C) = 1$ to obtain loopless optimal networks running the dynamics \Crefrange{eqn:periodic_dyn}{eqn:periodic_dyn_init_cond}. Our argument proceed as follows.

The matrix $C$ is symmetric by construction; thus if its rank is $1$ its eigenvalue decomposition is of the form $C = \sum_{i=1}^N \lambda_i x_i x_i^\top$, with all the eigenvalues equal to zero except one. We conventionally choose it to be $\lambda_1 = \sum_{v} C_{vv} > 0$, with a unit norm eigenvector $x_1$.  Defining $y := \sqrt{\lambda_1} x_1$ and substituting the eigendecomposition of $C$ in \Cref{eqn:periodic_rhs_def}, we get that $\bar{\Phi}_\Iedge$ is proportional to $\hat{\Flux}_\Iedge := ( \bar{\Cond}_\Iedge / \ell_\Iedge ) \sum_{\Inode} \Inc_{\Iedge \Inode} \hat{\Press}_\Inode$, with $\hat{\Press}_\Inode := \sum_u L^\dagger_{\Inode \Inodethree} (\bar{\Cond}) y_{u}$. In order to conclude, we need to show that $\hat{\Press}$ is a well-defined solution of \kl~law, $\sum_u L_{uv}(\bar{\Cond}) \hat{p}_\Inodethree = y_v$,  i.e., $y$ is a zero-sum vector \cite{klein1993resistance}.  This comes as a consequence of conservation of mass. Indeed, since for all times $\sum_v S_{v}(t) = 0$ holds, we have $\sum_v S_u(t) S_{v}(t) = 0, \, \forall \Inodethree \in \Nodes$.  Using \Cref{eqn:result_1_equation} and ignoring negligible terms of ${O}(\Delta)$, this yields $\sum_\Inode C_{\Inodethree \Inode} = 0, \, \forall \Inodethree \in \Nodes $. Finally, substituting the eigendecomposition of $C$ in this last relation gives $\sum_\Inode y_\Inodethree y_\Inode = 0, \, \forall \Inodethree \in \Nodes $. This is satisfied only if $\sum_v y_v = 0$, i.e., $y$ is a zero-sum vector. 
In this case, \Crefrange{eqn:periodic_dyn}{eqn:periodic_dyn_init_cond} correspond to the standard dynamics \Crefrange{eqn:fast_kirch_law}{eqn:fast_dyn_init_cond} with constant loads, which are $S(t) = y$, $\forall t \geq 0$, and we recover the well-known result that optimal networks are trees for $\gamma \leq 1$ \cite{maritan,bohn2007structure,kirkegaard2020optimal}.

Noticeably, if $\text{rank}(C) = 1$, it is possible to prove that the functional $\bar{\mathcal{L}}_\gamma  (\bar{\mu})$ proposed in \Cref{eqn:functional} is a well-defined Lyapunov functional. This means that for any $\bar{\Cond}(\tau)$ solution trajectory of \Crefrange{eqn:periodic_dyn}{eqn:periodic_dyn_init_cond}, we have $d\bar{\mathcal{L}}_\gamma(\bar{\Cond}(\tau))/d \tau \leq 0$, with stationarity achieved only by asymptotics of the dynamics.
Having established that $\hat{\Press}$ is a well-defined potential, we can write the Lyapunov functional as $\bar{\mathcal{L}}_\gamma  (\bar{\mu}) = (1/2) \sum_{v} \hat{p}_v (\bar{\Cond}) S_v + (1/2\gamma)\sum_{e} \ell_e \bar{\Cond}_\Iedge^\gamma$. This last expression is useful to conclude the proof, which follows that in Lonardi \textit{et al.} \cite{lonardi2021designing}.

\section{Derivation of \Cref{eqn:epression_b}}
\label{apx_sec:appendix_3}


\begin{figure*}[t]
 \centering
 \includegraphics[width=0.9\textwidth]{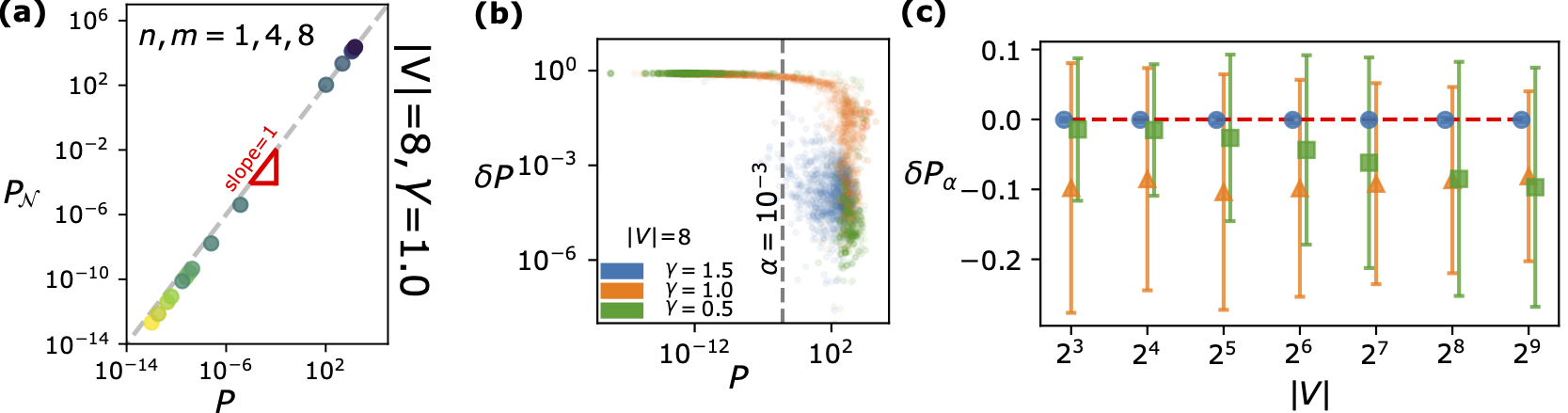}
 \caption{\label{fig:validation}Spectral density validation. (a) Plot of $P_\mathcal{N}$ versus $P$ for the example network of \Cref{fig:toymodel}. Each point corresponds to an edge; the color scale is that of  \Cref{fig:toymodel}a.  (b) Plot of $P$ versus $\delta P$. Each point corresponds to an edge, marker color denotes $\gamma = 0.5,1,1.5$. The dashed line is the cut-off used to build $\delta P_\alpha$. (c) Compatibility of $\delta P_\alpha$ with $\delta P_\alpha = 0$ for different networks' sizes. Markers and bars correspond to averages and standard deviations over $100$ random configurations of the problem, respectively.  The networks have been obtained pairing ten seeds for node coordinate generation and ten seeds for mass and conductivity initialization $\Cond_e \sim U(0,1)$.}
\end{figure*}


In order to discern the nature of the fast oscillating component $b_\Iedge(t)$ of the stabilized solutions, we need to investigate further the original dynamics \Crefrange{eqn:fast_kirch_law}{eqn:fast_dyn_init_cond}. From our numerical validation we observe that the fluxes start to oscillate around a constant value after a first stabilization time interval [see \Cref{fig:toymodel}a], analogously to the conductivities. This suggests the ansatz $\Flux_\Iedge(t) = \sum_{n_\Iedge \in \mathbb{Z}} \Flux_\Iedge^{n_\Iedge} \exp (\mathrm{i} \omega n_\Iedge t)$, $\forall \Iedge \in \Edges$, for all times $t$ sufficiently larger than $t_\text{STAB}$ and with the terms $\Flux_\Iedge^{n_\Iedge}$ amplitudes of the Fourier series decomposition. We argue that pairing this expression with \kl~law, i.e., $\sum_\Iedge \Inc_{\Inode \Iedge} \Flux_\Iedge(t) = \NetMass_\Inode(t)$, yields
\begin{align} \label{eqn:flux_series}
\Flux_\Iedge(t) = \sum_{n \in \mathcal{N}} \Flux_\Iedge^{n} \exp (\mathrm{i} \omega n t) \qquad \forall \Iedge \in \Edges \,,
\end{align}
with $\mathcal{N} := \{n_\Inode \}$ the set of Fourier modes of the loads injected in the network. Our argument is the following.

Assuming the ansatz $\Flux_\Iedge(t) = \sum_{n_\Iedge \in \mathbb{Z}} \Flux_\Iedge^{n_\Iedge} \exp (\mathrm{i} \omega n_\Iedge t)$, $\forall \Iedge \in \Edges$, we separate the contributions:
\begin{alignat}{2}
\label{eqnAPX:subflux_1}
F_e(t) &= \varphi_e(t) + \psi_e(t) \\
\label{eqnAPX:subflux_2}
\varphi_e(t) &= \sum_{n_\Iedge \in \mathcal{N} } \Flux_\Iedge^{n_\Iedge} \exp (\mathrm{i} \omega n_\Iedge t) \\
\label{eqnAPX:subflux_3}
\psi_e(t) &= \sum_{n_\Iedge \notin \mathcal{N} } \Flux_\Iedge^{n_\Iedge} \exp (\mathrm{i} \omega n_\Iedge t) \,.
\end{alignat}
Substituting \Crefrange{eqnAPX:subflux_1}{eqnAPX:subflux_3} in \kl~law returns the conditions:
\begin{alignat}{2}
\label{eqn:subkirck_1}
\sum_\Iedge B_{\Inode \Iedge} \varphi_e(t) &= S_v(t) \qquad  \\
\label{eqn:subkirck_2}
\sum_\Iedge B_{\Inode \Iedge} \psi_e(t) &= 0\,,
\end{alignat}
valid for all $ \Inode \in \Nodes$.
Now, in order to guarantee that the fluxes $\{ \varphi_e(t)$, $\psi_e(t) \}$ are well-defined, we suppose the existence of two time-dependent potentials $\alpha(t) = \{ \alpha_\Inode(t) \}$ and $\beta(t) = \{ \beta_\Inode(t) \}$. These are defined on the network nodes and such that for all $\Iedge \in \Edges $ we have
\begin{alignat}{2}
\label{eqn:subflux_1_def}
\varphi_e(t) &:= \frac{\Cond_\Iedge}{\ell_\Iedge} \sum_\Inode \Inc_{\Inode \Iedge} \alpha_\Inode(t) \qquad \\
\label{eqn:subflux_2_def}
\psi_e(t) &:=\frac{\Cond_\Iedge}{\ell_\Iedge} \sum_\Inode \Inc_{\Inode \Iedge} \beta_\Inode(t) \,.
\end{alignat}
Note that these definitions lead to $F_e(t)$ being a potential-based flux and yield $p_\Inode (t) = \alpha_\Inode(t) + \beta_\Inode(t), \forall \Inode \in \Nodes$. Substituting \Cref{eqn:subflux_1_def} and \Cref{eqn:subflux_2_def} in \Cref{eqn:subkirck_1} and \Cref{eqn:subkirck_2}, respectively, implies that $\psi_e(t) = 0$, $\forall \Iedge \in \Edges$, and for sufficiently large times. Hence, the only non zero terms in the Fourier decomposition of $\Flux_\Iedge(t)$ have modes in $\mathcal{N}$.

This result is particularly useful to describe the behavior of $\Cond(t)$ at large times. First, we recall that $\Cond_\Iedge(t) = \bar{\Cond}_\Iedge^\infty + b_\Iedge(t)$, $\forall \Iedge \in \Edges$, as discussed in \Cref{sec:characterization}. Moreover, we observe that in our numerical experiments [see \Cref{fig:toymodel}a] the size of the amplitude of the oscillatory term $b_\Iedge(t)$ is negligible in size with respect to $\bar{\Cond}_\Iedge^\infty$, unless $\Cond_\Iedge(t)$ decays to zeros. This allows us to approximate \Cref{eqn:fast_dyn} as
\begin{align} \label{eqn:approx_dyn}
\f{d \Cond_\Iedge(t)}{dt} \simeq \f{\Flux_\Iedge^2 (t)}{ (\bar{\Cond}^{\infty}_\Iedge)^\gamma } - \Cond_\Iedge(t) \qquad \forall \Iedge \in \Edges \,.
\end{align}

Finally, substituting \Cref{eqn:flux_series} in \Cref{eqn:approx_dyn}, we get the desired results, i.e., the main oscillatory modes of the conductivities, hence of $b_\Iedge(t)$, are resonant with the squared fluxes. Thus we obtain \Cref{eqn:epression_b}.

\subsection{Validation on synthetic networks}

We test these expressions numerically on networks generated as described in \Cref{sec:characterization}. We compute $P_\Iedge := \int_{\mathbb{R}} |\mathcal{F}[b_\Iedge](f)|^2 \,  d f$, the total spectral density of the oscillatory components $b_\Iedge(t)$, after the conductivities $\Cond(t)$ stabilize. Here $\mathcal{F}[\cdot](f)$ is the Fourier transform operator. Additionally, we calculate $P_{\mathcal{N}}$, obtained summing the atomic contributions of the spectral density on the modes $k \in \mathcal{K} := \{k \; \text{s.t.} \; k = n + m$, for $n,m \in \mathcal{N} \}$. Namely, $P_{\mathcal{N},\Iedge} := \sum_{k \in \mathcal{K}} \int_{\mathbb{R}} |\mathcal{F}[b_\Iedge](f)|^2 \delta (f - k) \,  d f, \forall \Iedge \in \Edges$. 

From \Cref{eqn:epression_b} we expect to have most of the spectral density of $b_\Iedge(t)$ concentrated on the modes in $\mathcal{K}$, i.e., the ratio $P_\Iedge/P_{\mathcal{N},\Iedge}$ should be close to $1$ for each edge. In \Cref{fig:validation}a we plot $P = \{P_\Iedge\}$ versus $P_\mathcal{N} = \{ P_{\mathcal{N},\Iedge} \} $ for the example network considered in \Cref{fig:toymodel}.
The plot supports \Cref{eqn:epression_b}; indeed, the element-wise ratio $P/P_\mathcal{N}$ is close to $1$ for all points (each correspondent to a different edge) with a slight deviation only for small (thus negligible) values of the conductivities.

We further validate this result on an additional synthetic example network. We construct the Delaunay networks described in \Cref{sec:characterization} considering $100$ combinations of seeds for the nodes' positions and for the random input loads. Then, we compute the spectral densities $P$ and plot them against $\delta P$, with entries $\delta P_\Iedge := (P_{\mathcal{N},\Iedge} - P_\Iedge)/P_\Iedge$. We show in \Cref{fig:validation}b results for $\gamma = 0.5,1,1.5$ on $100$ random graphs of size $|V|=8$. Here we clearly see that $\delta P$ are negligible for any edge with $P$ larger than a threshold $\alpha$ (in our experiments we set $\alpha=10^{-3}$), further supporting the result in \Cref{fig:validation}a.

It is worth mentioning how the points cluster in different regions of the plot for different values of $\gamma$. The green points,  corresponding to $\gamma = 0.5$, are divided into two clusters: one around $P$ small and $\delta_P=1$ and another with $P$ large and $\delta P$ negligible. This reflect the tendency of $\gamma < 0.5$ to aggregate fluxes on few edges. The blue points, corresponding to $\gamma = 1.5$, are instead concentrated around a region with $P$ large and $\delta P$ small, since in this case fluxes are distributed on more edges. Finally, the orange points, corresponding to $\gamma = 1$, represent a transition between the two cases and are located in a cluster placed in between the other two. This result is consistent with the behavior of $\gamma$ mentioned in \Cref{sec:dyn}.

We test the scalability of our result by running the same validation just described, but increasing the graphs sizes. We plot our results in \Cref{fig:validation}c. Here we show the compatibility of $\delta P_{\alpha} := \sum_{\Iedge} \delta P_\Iedge \mathbb{I}({ \delta P_\Iedge > \alpha}) / E' $ with zero.  Here $\mathbb{I}(\cdot)$ is the indicator function, and $E'$ the number of the edges that do not get trimmed by $\alpha$. We see that all values attain values close to $\delta P_\alpha = 0$ and all errorbars (expressing standard deviations over 100 random graph realizations) are always intersecting the line highlighting $\delta P_\alpha = 0$. The decreasing trend of $\delta P_\alpha$ for $\gamma = 0.5$ can be attributed to the fact that we fixed the cut off threshold $\alpha$ a priori and thus we do not have a precise trim for $P$ for larger networks.

\section{Harmonic oscillator conditions}
\label{apx_sec:appendix_4}

We already established that if $\text{rank}(C) = 1$, then there exists a zero-sum vector $y$ such that $C_{\Inodethree \Inode} = y_\Inodethree y_\Inode, \forall \Inodethree, \Inode \in \Nodes$ (see \Cref{apx_sec:appendix_2}). Inspecting the diagonal elements of $C$, it is immediate to get $y_\Inode = \pm \sqrt{C_{\Inode \Inode}}, \forall \Inode \in \Nodes$. Here the choice of the plus or minus sign is constrained among $2^{\Nnode}$ possibilities, to those for which $\sum_\Inode y_\Inode = 0$ holds. 
The right-to-left implication comes naturally from the definition of $C$. Namely, if we suppose that $C_{\Inodethree \Inode} = y_\Inodethree y_\Inode, \forall \Inodethree, \Inode \in \Nodes$,  we are imposing that all the columns of $C$ are scalar multipliers of each other, i.e. , $\text{rank}(C) = 1$.

Substituting \Cref{eqn:importan_rel_harmonic} in $\nu_\Inode = \lambda \nu_\Inodethree$ leads to
\begin{align}
\label{eqn:apx_detailed_harmonic}
A^i_{\Inode} \exp(\mathrm{i} \phi_\Inode^i ) \delta_{n_\Inode n_\Inode^i} = \lambda A^i_{\Inodethree} \exp(\mathrm{i} \phi_\Inodethree^i ) \delta_{n_\Inodethree n_\Inodethree^i} \,,
\end{align}
which needs to be satisfied for each pair of $n_\Inodethree, n_\Inode \in \mathbb{N}$.  This is valid if the phases are such that $\phi_\Inode^i = \phi_\Inodethree^i + k\pi, \, k \in \mathbb{Z}$, i.e., condition (i) in \Cref{ssec:conditions_closed_form} holds. Substituting this last equality in \Cref{eqn:apx_detailed_harmonic}, we get
\begin{align}
\label{eqn:apx_detailed_harmonic_1}
A^i_{\Inode} \exp(\mathrm{i} \phi_\Inode^i ) \delta_{n_\Inode n_\Inode^i} &= \lambda A^i_{\Inodethree} \exp(\mathrm{i} \phi_\Inode^i )(-1)^k \delta_{n_\Inodethree n_\Inodethree^i} \\
A^i_{\Inode} \delta_{n_\Inode n_\Inode^i} &= \lambda A^i_{\Inodethree} (-1)^k \delta_{n_\Inodethree n_\Inodethree^i} \,,
\end{align}
which is precisely (ii) in \Cref{ssec:conditions_closed_form}.
In conclusion, fixing the input loads in such a way that (i) and (ii) holds lead to $\text{rank}(C)=1$, which is sufficient to get optimal tree topologies, as shown in \Cref{apx_sec:appendix_2}.

\bibliographystyle{apsrev4-1}
\bibliography{bibliography}

\end{document}